# Spin splitting and even-odd effects in carbon nanotubes

David H. Cobden†, Marc Bockrath, and Paul L. McEuen
*Department of Physics, University of California and Materials Science Division,*
*Lawrence Berkeley National Laboratory, Berkeley, California, 94720*

Andrew G. Rinzler and Richard E. Smalley
*Center for Nanoscale Science and Technology, Rice Quantum Institute, and*
*Department of Chemistry and Physics, MS-100, Rice University, P.O. Box 1892, Houston, TX 77251*

(March 12, 1998)

The level spectrum of a single-walled carbon nanotube rope, studied by transport spectroscopy, shows Zeeman splitting in a magnetic field parallel to the tube axis. The pattern of splittings implies that the spin of the ground state alternates by ½ as consecutive electrons are added. Other aspects of the Coulomb blockade characteristics, including the current-voltage traces and peak heights, also show corresponding even-odd effects.

The spin state of small multi-electron systems is an important testing ground for our understanding of interacting quantum systems. For $N$ non-interacting electrons in non-degenerate levels with spin, the single-particle states are occupied in order of energy, leading to a total spin $S = 0$ for even $N$ and $S = 1/2$ for odd $N$. Coulomb interactions among the electrons can alter this behavior, however. In atoms, for example, the exchange interaction among electrons in a shell leads to Hund's rule and a spin-polarized ground state for a partially filled shell. Recently, attention has been focused on similar questions in quantum dots. In small 3D metallic dots, Zeeman splitting consistent with an alternation between $S = 0$ and $1/2$ was found [1]. This may be understood within the constant interaction (CI) model [2], where the energy for adding an electron is the non-interacting level spacing $\Delta E$ plus a constant charging energy $U$. On the other hand, in two-dimensional dots evidence for spin polarization in the ground state has been found in recent experiments on both high symmetry [3] and low symmetry dots [4], requiring explanations beyond the CI model.

Of considerable interest is the situation in 1D, where Coulomb interactions are predicted to profoundly influence the properties of the system [5]. Here exact theoretical results are available for many model systems. For instance, for electrons in a box in strictly one dimension (1D), Lieb and Mattis [6] proved that in spite of interactions the ground state has the lowest possible spin. In real systems, however, a variety of factors, such as finite transverse dimensions, multiple 1D subbands, and spin-orbit coupling, may lead to a spin-polarized ground state.

Here we present measurements of the spin state of single-walled carbon nanotubes, a novel quasi-1D conductor where the current is carried by two 1D subbands [7]. It has recently been shown experimentally [8,9] that when contacts are attached, these nanotubes behave as quasi-1D quantum dots. Here we concentrate on a very short (~200 nm) nanotube dot with a correspondingly large level spacing. To study the spin state, we apply a magnetic field along the axis of the nanotube and examine the Zeeman effects in the transport spectrum. From the pattern of the spin splitting, we conclude that as successive electrons are added the ground state spin oscillates between $S_0$ and $S_0+1/2$, where $S_0$ is most likely zero. This results in an even/odd nature of the Coulomb peaks which is also manifested in the asymmetry of the current-voltage characteristics and the peak height. It may also be reflected in the excited state spectrum.

The devices are made [9] by depositing single-walled nanotubes [10] from a suspension in dichloroethane onto 1-μm thick $SiO_2$. The degenerately doped silicon substrate is used as a gate electrode. A single rope is located relative to prefabricated gold alignment marks using an atomic force microscope (AFM). Chromium-gold contacts are then deposited on top using 20 keV electron beam lithography. An AFM image of a 5-nm diameter rope (consisting of about a dozen tubes) with six contacts is shown in the inset to Figure 1. Leads labeled s (source), d (drain) and $V_g$ (gate) are drawn in to indicate the typical measurement configuration.

Figure 1 shows the linear-response two-terminal conductance, $G$, versus gate voltage, $V_g$, at magnetic field $B = 0$ and temperature $T = 100$ mK. It exhibits a series of sharp Coulomb blockade oscillations [2,8,9] that occur each time an electron is added to the nanotube dot. For $T <\sim 10$ K all the peaks have the same width, proportional to $T$ [9], and a $T$-independent area, indicating that the level spacing $\Delta E$ is $\gg k_B T$ and that transport is through a single quantum level. We deduce that the dot electrostatic potential $V_{dot}$ is linearly related to $V_g$, with a coefficient $\alpha \equiv dV_{dot}/dV_g = 0.09$.

Figure 2 (a) is a greyscale plot of the differential conductance $dI/dV$ as a function of $V$ and $V_g$ at $B = 0$. Dark lines here are loci of peaks in $dI/dV$. Crosses P0 and P1 are formed by the identically labeled Coulomb peaks in Figure 1. The interpretation of such a plot in the CI model is well known [3]. Each line is produced by the alignment of a quantized energy level in the dot with the Fermi level in a contact. From the spacing of the lines we infer a typical level spacing $\Delta E \sim 5$ meV, and from the average Coulomb peak spacing we obtain a charging energy $U \sim 25$ meV. These values are consistent with expectations based on previous measurements [8,9] for a 100-200 nm length of tube. Thus we find as before [9] that the portion of nanotube rope forming the dot appears roughly equal in length to the distance between the contacts (nominally 200 nm.)

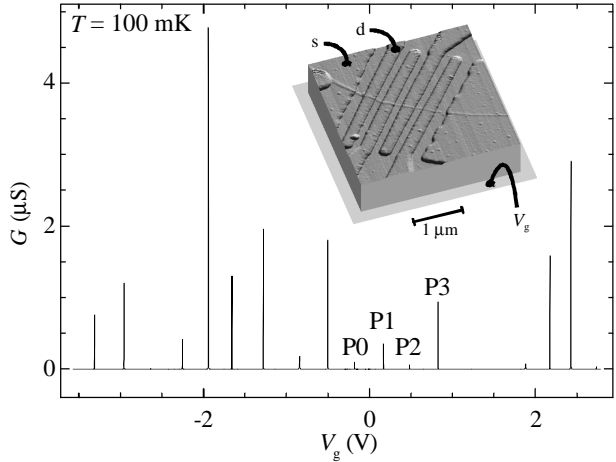

FIG. 1. (a) Conductance $G$ of a nanotube rope vs gate voltage $V_g$. Inset: AFM image of a device with schematic wires added.

Figure 2 (b) shows the results of the same measurement at $B = 5$ T. Most of the lines observed at $B = 0$ have split into parallel pairs. The splitting is linearly proportional to B. This can be seen in Figure 2 (c), where the relative positions of the peaks in $dI/dV$ at $V = -7$ mV (dotted line in Figure 2 (a)) are plotted as a function of $B$. One group of peaks (denoted by open symbols) moves downwards in $V_g$ relative to the other (solid symbols) by an amount proportional to $B$. Note that not all the lines at $B = 0$ split. Over a series of ten consecutive crosses in the range $-2$ V $< V_g <$ $+1$ V [11], the following pattern emerges: on alternate peaks, (P0, P2, etc.,) the leftmost lines in the cross (such as T) do not split, while on the other peaks, (P1, P3, etc.,) the rightmost lines (such as Z) do not split.

These measurements can be used to obtain information about the ground-state spin $S_N$ of the dot with $N$ electrons, as we now discuss. The analysis is based on the following spin selection rules: since the tunneling electron carries spin 1/2, both the total spin, $S$, and its component along the magnetic field axis, $S_z$, must change by $\pm 1/2$ for observable transitions [12].

The energy required for a tunneling process is the energy difference between the $N$- and $(N+1)$-electron states. In the absence of orbital effects [13], this depends on $B$ only through the Zeeman term $-g\mu_B B \Delta S_z$, where $g$ is the electronic g-factor, $\Delta S_z$ is the change in $S_z$ and $\mu_B$ is the Bohr magneton. In Figure 2 (c) we therefore associate the open-symbol transitions with $\Delta S_z = +1/2$ and the closed-symbol transitions with $\Delta S_z = -1/2$. Fitting their separation to $g\mu_B B/\alpha$ yields $g = 2.04 \pm 0.05$, which is consistent with $g = 2.0$ for graphite and with the value $g = 1.9 \pm 0.2$ obtained previously for a single excited state in a nanotube [8].

From the pattern of splittings of the lowest-energy transitions (the edges of the crosses in Figure 2 (a)) one can deduce the change in ground-state spin, $\Delta S = S_{N+1} - S_N = \pm 1/2$, across each Coulomb peak. The reason is as follows [1]. First consider an electron tunneling into the $N$-electron ground state in a magnetic field, where initially the total spin is aligned with the field, so that $S_z = -S_N$. For the case $\Delta S = +1/2$, after tunneling $S_z$ may be either $-S_N - 1/2$ or $-S_N + 1/2$. The corresponding line therefore splits with $B$. However, for the case $\Delta S = -1/2$, only $S_z = -S + 1/2$ is possible for the final state, because of the requirement $|S_z| \le S_{N+1} = S - 1/2$. The corresponding line therefore does not split with $B$. A similar argument for an electron tunneling out of the $N+1$ ground state shows that if $\Delta S = -1/2$ the line splits, while if $\Delta S = +1/2$ it does not. To summarize: if $\Delta S = +1/2$ for a Coulomb peak, the lines on the right edge of the cross do not split, while if $\Delta S = -1/2$ the lines on the left edge do not split.

This general result is also predicted by the CI model, as indicated in Figure 3. If $N$ is even, $S_N = 0$, and the

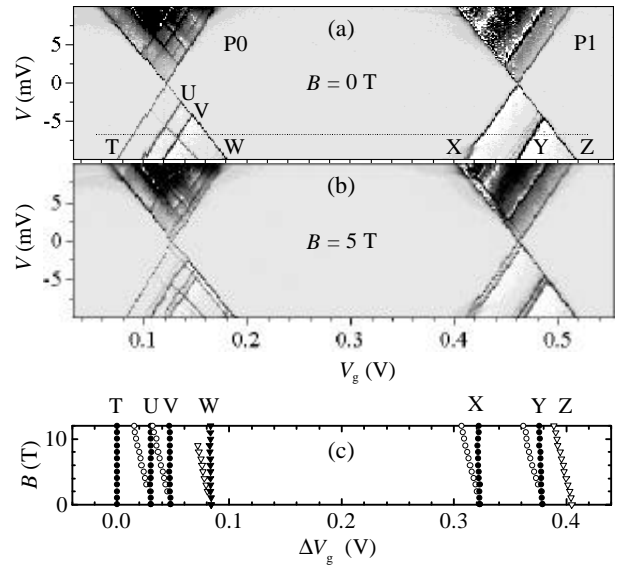

FIG. 2. (a) Greyscale plot of the differential conductance $dI/dV$ of Coulomb peaks P0 and P1 at $B = 0$ (darker = more positive $dI/dV$.) (b) Same as (a) but at $B = 5$ T. (c) $B$-dependence of the relative positions of the peak in $dI/dV$ labeled T-Z in (a), at a bias of $V = -7$ mV as indicated by the dashed line in (a). On the x-axis we plot $\Delta V_g = V_g - V_g^T$, where $V_g^T$ is the position of peak T, to remove unreproducible temporal drift of the characteristics along the $V_g$ - axis.

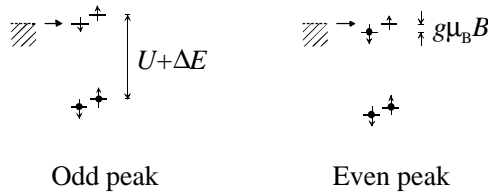

Odd peak          Even peak

FIG. 3. Explanation of splitting pattern within the CB model. The lowest-energy transition splits for an odd peak (N changes from even to odd) but not for an even peak (N changes from odd.to even.)

next electron can be added to either spin-up or spin-down state of the next orbital level (left sketch), resulting in $S_{N+1} = 1/2$. On the other hand, if $N$ is odd, $S_N = 1/2$ and the next electron can only be added to the one empty spin state of that level (right sketch), resulting in $S_{N+1} = 0$. A corresponding story can be told for removing an electron. The predicted pattern of splittings is the same as in the previous paragraph, but with the additional implication that $N$ is even if $\Delta S = +1/2$ and odd if $\Delta S = -1/2$.

Comparing the above predictions with Figure 2, we find that $\Delta S = +1/2$ for peak P0 and $\Delta S = -1/2$ for peak P1. Since the pattern of splitting alternates between the two types over ten Coulomb peaks, we deduce that $S_N$ oscillates between some value $S_0$ and $S_0+1/2$ as ten successive electrons are added. We cannot rule out the possibility that $S_0$ is finite. However, since polarization of a system is usually related to states near the Fermi level, and in this system we see the spin alternating as these states are filled, it is most likely that that $S_0 = 0$, as in the CI model. If this is the case, the behavior is consistent with the prediction of Ref. [6] for 1D electrons: the ground state spin alternates between 0 and 1/2. This is our principal result. We subsequently describe Coulomb peaks where $N$ changes from odd to even (P0, P2, etc.) as *even* peaks, because the added electron is even. Peaks P1, P3, etc. we call *odd* peaks, because the added electron is odd. This is indicated in Figure 3.

The alternating spin of the ground state should also be reflected in the *I-V* characteristics at zero magnetic field, if the source and drain contacts have different tunnel resistances. If, for instance, the source contact dominates the resistance, the magnitude of the current $I_-$ at negative source bias $V$ is determined by transitions from the $N$ to the $N+1$ electron ground state, as long as the bias is less than the level spacing. On the other hand, the current $I_+$ at positive $V$ is determined by transitions from the $N+1$ to the $N$ electron ground state. The ratio $\beta = I_+/I_-$ therefore reflects the differences caused by the spin selection rules in these two situations. This can easily be understood in the CI model, as illustrated for an even peak ($\Delta S = -1/2$) in Figure 4 (a). For negative $V$ (left sketch) an electron tunneling in from the source can only go into one available spin state. On the other hand, for positive $V$ (right sketch), either of two electrons can tunnel out. The current is therefore larger for positive $V$. An elementary calculation gives $\beta = (G_s+2G_d)/(2G_s+G_d)$, where $G_s$ and $G_d$ are the source and drain barrier conductances respectively. For $G_s < G_d$, this predicts $1 < \beta < 2$. In contrast, for an odd peak ($\Delta S = +1/2$), the inverse ratio is found, and $1/2 < \beta < 1$ is predicted.

The solid line in figure 4 (b) is the *I-V* characteristic measured at the center of peak P0. Near $V = 0$, the *I-V* is ohmic, but for $|V| >\sim 0.5$ mV the current saturates into a slowly varying form. The saturation current is larger for positive than for negative $V$. Moreover, if the same data is plotted (dashed line) with the current scaled by a factor $-\beta$, where $\beta = 1.57$, the *I-V*'s in the two bias directions can be brought onto the same interpolated curve (dotted line.) For each peak an appropriate value of $\beta$ can be chosen to achieve a similar matching. The results are plotted in the top panel of Figure 4 (c). We find that $1 < \beta < 2$ for P0 and P2, while $1/2 < \beta < 1$ for P1 and P3. Comparing these values with the predictions for $\beta = I_+/I_-$ in the previous paragraph, we see that they are perfectly consistent with our assignments of $\Delta S = +1/2$ or $-1/2$ from the Zeeman splitting [14].

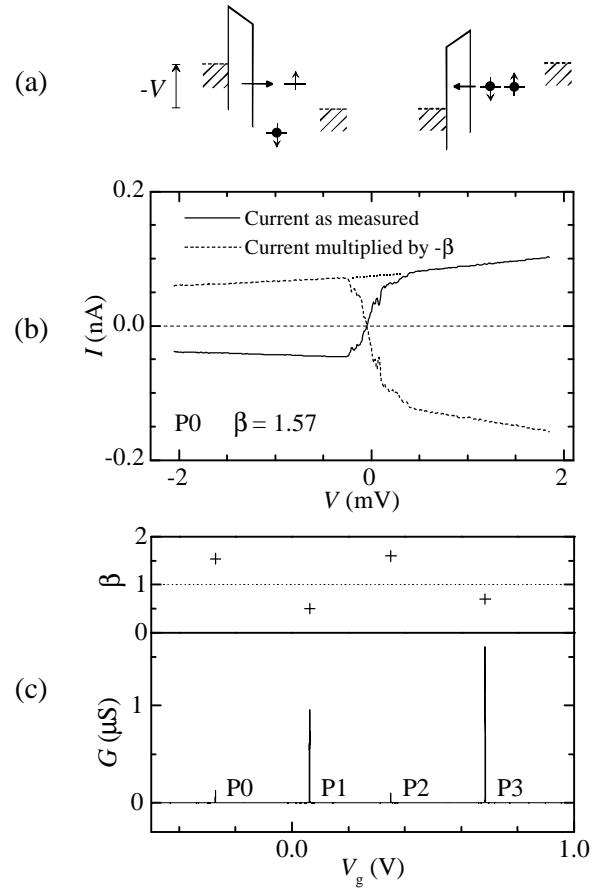

FIG. 4. (a) Current flow at high bias in the CI model. Only the larger barrier, between source and dot, is drawn. (b) Solid line: *I-V* measured at the center of peak P0 in Figure 1. Dotted line: the same trace with $I$ multiplied by $-\beta = -1.57$. (c) Lower: expanded view of the peaks P0 - P3. Upper: measured values of $\beta$ for these peaks. The oscillating value of $\beta$ implies that successive electrons are added with opposite spin directions (see text).

We have seen from the Zeeman splitting and the *I-V* characteristics that the ground state spin behaves as is predicted by the CI model. However, this implies not that effects such as exchange are small, but only that they do not change the spin of the *N*-electron ground state of the system. Exchange might for instance be manifested in the excited state spectra, where one would anticipate a difference between even and odd peaks. For odd peaks, the added electron simply goes into higher unoccupied orbital levels, giving rise to a single-particle spectrum. For even peaks, however, the added electron can form singlet and triplet states with the original unpaired electron, leading to exchange splitting. A singlet-triplet splitting has indeed been seen in the excitation spectra of semiconductor dots [15]. We observe indications of this predicted behavior in peaks P0-P3. The lowest excited states visible at negative *V* on even peaks in each case form a pair (such as lines U and V on peak P0 in Figure 2(a)), while those on odd peaks do not (such as line Y on P1). This will be investigated further in future work.

A contradiction with the CI model is also seen in the peak heights. These are predicted to be identical for a pair of peaks arising from a single orbital level [16]. However, we find that the odd peaks tend to be considerably larger than the even peaks, as apparent in Figure 4 (c). This behavior is not understood and deserves further investigation.

In summary, our transport measurements of a short nanotube quantum dot show that the ground state of this 1D electronic system alternates between $S = 0$ and $S = 1/2$. A variety of even-odd effects are seen in the addition spectrum, some of which, such as an alternation of the peak heights, require explanations beyond the simple Coulomb blockade picture.

We thank Marvin Cohen, Cees Dekker, Noah Kubow, Dung-Hai Lee, Steven Louie, Jia Lu, Boris Muzykantskii, and Sander Tans for helpful discussions. The work at LBNL was supported by the Office of Naval Research, Order No. N00014-95-F-0099 through the U.S. Department of Energy under Contract No. DE-AC03-76SF00098, and by the Packard Foundation. The work at Rice was funded in part by the National Science Foundation and the Robert A. Welch Foundation.

---

†Present address: Oersted Laboratory, Universitetsparken 5, DK-2100 Copenhagen Oe., Denmark.
E-mail: dcobden@physics.berkeley.edu

[1] D. C. Ralph, C. T. Black, and M. Tinkham, Phys. Rev. Lett. **74**, 3241 (1995); Phys. Rev. Lett. **78**, 4087 (1997).
[2] L. P. Kouwenhoven *et al.*, in *Mesoscopic Electron Transport*, Eds. L. P. Kouwenhoven, G. Schon and L. L. Sohn (Kluwer, 1997).
[3] S. Tarucha *et al.*, Phys. Rev. Lett. **77**, 3613 (1996).
[4] D. R. Stewart *et al.*, Science **278**, 1784 (1998).
[5] See, e.g., *The Many-Body Problem*, Ed. D. C. Mattis (World Scientific, Singapore, 1993).
[6] Elliot Lieb and Daniel Mattis, Phys. Rev. **125**, 164 (1962).
[7] N. Hamada, S. Sawada, and A. Oshiyama, Phys. Rev. Lett. **68**, 1579 (1992); R. Saito *et al.*, Appl. Phys. Lett. **60**, 2204 (1992).
[8] Sander S. Tans *et al.*, Nature **386**, 474 (1997).
[9] Marc Bockrath *et al.*, Science **275**, 1922 (1997).
[10] Andreas Thess *et al.*, Science **273**, 483 (1996).
[11] Outside this range of $V_g$ the characteristics are complicated by charging of another dot (probably another section of nanotube.)
[12] D. Weinmann, W. Hausler, and B. Kramer, Phys. Rev. Lett. **74**, 984 (1995).
[13] The only expected orbital effect of an axial magnetic field results from the Aharonov-Bohm phase due to the flux φ through the tube [17], which may shift the levels by up to ~ 2.5 meV at 12 T. In our data the non-Zeeman energy shifts of the transitions at this field are less than 100 μeV This can be accounted for if all levels are shifted equally, as expected if the Fermi level is displaced away from the band-crossing points by, for example, charge transfer from the metal contacts [18].
[14] From a detailed study of the *I-V*'s in this range of $V_g$ we can deduce that $G_s < G_d$. The gradual increase of |*I*| as *V* becomes more positive in Figure 4 (b) is explained by electric-field lowering of the dominating source barrier, as indicated in Figure 4 (a).
[15] L. P. Kouwenhoven *et al.*, Science **278**, 1788 (1997).
[16] C. W. J. Beenakker, Phys. Rev. B **44**, 1646 (1991).
[17] H. Ajiki and T. Ando, J. Phys. Soc. Jap. **62**, 1255 (1993); C. L. Kane and E. J. Mele, Phys. Rev. Lett. **78**, 1932 (1997).
[18] Jeroen W. G. Wildoer *et al.*, Nature **391**, 59 (1998).